\documentclass[preprint,preprintnumbers,amsmath,amssymb]{revtex4}
\usepackage[latin1]{inputenc}
\usepackage{amsmath}
\usepackage{amsfonts}
\usepackage{amssymb}
\usepackage{natbib}
\usepackage{graphicx}
\usepackage{bm}

\begin{document}
%\preprint{APS/123-QED}
%\begin{figure}[b]
%\includegraphics[width=0.4\textwidth]{figure1.ps}
%\caption{Schematic of the Sch\"affler diagram. Concentrations are
%given in weight percent. The numbers represent a few commercially
%available duplex steel alloys. } \label{fig1}
%\end{figure}
\title{Initial oxidation of Fe-Al and Fe-Cr-Al alloys:  Cr as an alumina booster}
\author{M.H. Heinonen} 
\author{K. Kokko}  
\affiliation{Department of Physics and Astronomy, University of Turku, FI-20014 Turku, Finland}
\affiliation{Turku University Centre for Materials and Surfaces (MatSurf), Turku, Finland}
\author{M.P.J. Punkkinen} 
\affiliation{Department of Physics and Astronomy, University of Turku, FI-20014 Turku, Finland}
\affiliation{Turku University Centre for Materials and Surfaces (MatSurf), Turku, Finland}
\affiliation{Department of Physics and Materials Science, Uppsala University, SE-75121 Uppsala, Sweden} 
\author{E. Nurmi}\email{eero.nurmi@utu.fi} 
\affiliation{Department of Physics and Astronomy, University of Turku, FI-20014 Turku, Finland}
\affiliation{Turku University Centre for Materials and Surfaces (MatSurf), Turku, Finland}
\affiliation{Graduate School of Materials Research, Turku, Finland}
\author{J. Koll\'ar} 
\affiliation{Research Institute for Solid State Physics and Optics, Budapest
H-1525, P.O. Box 49, Hungary} 
\author{L. Vitos} 
\affiliation{Department of Physics and Materials Science, Uppsala University, SE-75121 Uppsala, Sweden} 
\affiliation{Research Institute for Solid State Physics and Optics, Budapest H-1525, P.O. Box 49, Hungary} 
\affiliation{Applied Materials Physics, Department of Materials Science and Engineering, Royal Institute of Technology, SE-10044 Stockholm, Sweden}
\date{20 Janyary 2011}
\begin{abstract}
The boosting effect of Cr on the growth of the protective alumina scale on Fe-Al alloys is investigated by x-ray photoelectron spectroscopy. Using low oxygen pressure the surface chemistry of the alloys is monitored starting from the first moments of oxidation. Chromium effect on the Fe-Al surface-bulk exchange is clearly detected by analyzing the measured surface concentrations within the atomic concentration models. Previous {\em ab initio} calculations agree well with the present experiments. 

\end{abstract}
\maketitle
%\section{Introduction} 
Fe-Al alloys are frequently used as base materials in high-temperature applications due to their good corrosion resistance which is based on a highly stable and protective Al-oxide scale. \cite{case_1953,khanna_2002,khanna_2005} Due to the brittleness of the high Al alloys there is a need to reduce the Al concentration in the bulk without affecting the Al content at the surface. It is known that adding Cr improves the oxidation resistance of Fe-Al alloys. \cite{tomaszewicz_1978,prescott_1992,stott_1995,brady_2000,gurrappa_2000,babu_2001,lee_2003,Hou_2003,wright_2004,nychka_2005,zhang_2006,asteman_2008,niu_2008} To achieve a better technical control to this basically unexplained boosting phenomenon a detailed understanding of its basic atomic mechanisms is indispensable. 

In our previous investigations ambient conditions oxidation was used. Then, by successive sputtering and Auger electron spectroscopy (AES), it was possible to construct a time reversed picture of the oxidation. \cite{airiskallio_2010a,airiskallio_2010b} In the present work we address the direct observation of the first moments of the oxidation of Fe13Al and Fe10Cr10Al (at. \% used in alloy formulas) by using x-ray photoelectron spectroscopy (XPS) and low oxygen pressure. By combining our previous AES and present XPS results, complementary in respect to time scales of oxidation, as well as our previous density functional study more detailed description of the oxidation of Fe-Al based alloys can be accomplished.

%\section{Methods}
The alloys were made by induction melting under argon flow from elemental components of purity better than 99.99\%. Plates of about 3 mm thick were cut from the ingots and then ground with SiC-paper down to 1000 grit. Final polishing was done with 1 $\mu$m diamond suspension. The bulk concentrations are shown in Table \ref{t_bulk}. 
\begin{table}[h]
\centering
\caption{\label{t_bulk}Bulk concentrations determined by energy-dispersive x-ray spectroscopy.}
\begin{tabular}{lccc} 
\hline
\hline
alloy& Fe (at. \%)& Al(at. \%)& Cr (at. \%)\\
\hline
Fe13Al    & 87.4& 12.6& -\\
Fe10Cr10Al& 79.9&  9.8& 10.3\\
\hline
\hline
\end{tabular}
\end{table} 

Measurements were done with a PHI 5400 spectrometer using twin anode Al K$\alpha$ excitation. The pressure in the analyser chamber was $5\times 10^{-10}$ Torr during the measurements. The energy scale was calibrated with Ag 3d$_{5/2}$ (368.2 eV) and Cu 2p$_{3/2}$ (932.6 eV) lines. Samples were cleaned by repeated cycles of sputtering with 3 keV Ar$^{\rm +}$  ions and annealing at 700 $^{\rm o}$C. Sample heating was done by a resistive heater from the rear of the sample. Temperature was measured with a thermocouple in contact with the sample. Cleaning cycle was repeated until no contamination peaks were seen in the XPS spectra. Oxidation of the samples was carried out in a preparation chamber connected to the spectrometer. Oxygen pressure during oxidation was $1\times 10^{-7}$ Torr up to 100 L and at higher doses $1\times 10^{-6}$ Torr. Atomic concentrations were calculated from the peak areas after Shirley-type background subtraction using sensitivity factors given by the manufacturer of the spectrometer.

\begin{figure}[h]
\includegraphics[width=0.27\textwidth,angle=0]{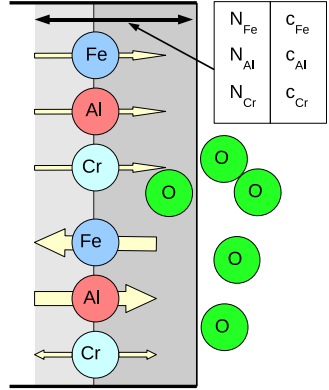}
\caption{The shaded background shows the model system. Darker shading represents the region of direct influence of oxygen. The upper three atomic types illustrate a typical average atomic migration coming with the FCR model. The lower three atomic types represent a typical VCR model case. 
%The quantities $N_{\rm X}$ and $c_{\rm X}$, Eqns. (\ref{e_numb}) and (\ref{e_con}), are the number of atoms and concentrations in the gray region.}  
\label{f_schema}}
\end{figure}

To get more detailed information of the early stage of oxidation the experimental data was analyzed using parameterized models. Our model system (Fig.\ \ref{f_schema}) consists of surface atomic layers up to the probe depth of the XPS. Two oxidation models were introduced and their predictions were compared with the XPS measurements. First model assumes the ratios of the number of different type metal atoms in the model system to be independent on the oxygen dose. This fixed cations ratio (FCR) model describes e.g.\ cases where cationic diffusion between the surface system and the rest of the alloy does not depend on cationic type or the size of the system changes due to the variations of the XPS probe depth.  Another model takes into account atomic processes capable of changing the relative abundances of the metal atoms in the model system. This variable cations ratio (VCR) model describes e.g.\ cases where diffusion is significant and metal atoms of different type have different diffusion characteristics. 

Since we are interested in the very first moments of oxidation it is advantageous to parameterize the number of metal atoms in the model system to be proportional to their initial values 
\begin{equation} 
N_{\rm X}(\theta)=(1+k_{\rm X}(\theta))N_{\rm X}(0),\quad k_{\rm X}(0)=0 \label{e_numb}
\end{equation} 
where $\theta$ is the oxygen dose, $N_{\rm X}$ is the number of atoms of type X = Fe, Cr, Al, and $k_{\rm X}$ are $\theta$-dependent parameters.
%\begin{equation} 
%k_{\rm Fe}(0)=k_{\rm Cr}(0)=k_{\rm Al}(0)=0. \label{e_init}
%\end{equation} 
The FCR model corresponds to cases where the $k_{\rm X}$ parameters are identical $k_{\rm Fe}(\theta)\equiv k_{\rm Cr}(\theta)\equiv k_{\rm Al}(\theta)$ whereas all other cases come with the VCR model.

\begin{figure}[h]
\includegraphics[width=0.32\textwidth,angle=270]{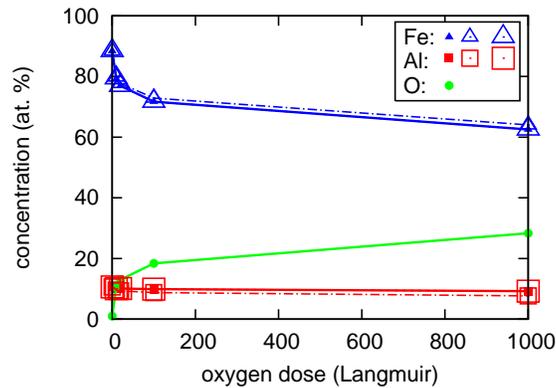}
\caption{Atomic concentrations of the surface of Fe13Al at room temperature as a function of oxygen dose. Solid curve with filled symbols: experiment; chain curve with small open symbols: FCR model; and dashed curve with large open symbols: VCR model.} \label{fig1}
\end{figure}

\begin{figure}[h]
\includegraphics[width=0.32\textwidth,angle=270]{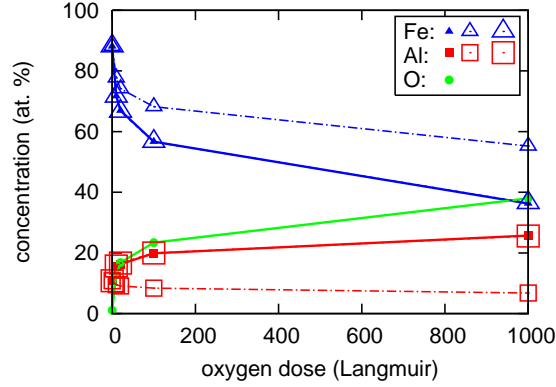}
\caption{The surface of Fe13Al at 700 $^{\rm o}$C, curves and symbols as explained in Fig.~\ref{fig1}.} \label{fig2}
\end{figure}

\begin{figure}[h]
\includegraphics[width=0.32\textwidth,angle=270]{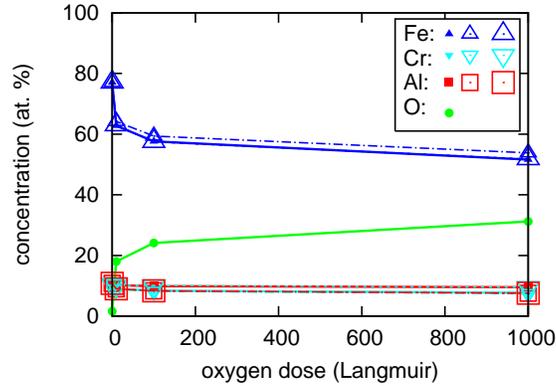}
\caption{The surface of Fe10Cr10Al at room temperature, curves and symbols as explained in Fig.~\ref{fig1}.} \label{fig3}
\end{figure}

\begin{figure}[h]
\includegraphics[width=0.32\textwidth,angle=270]{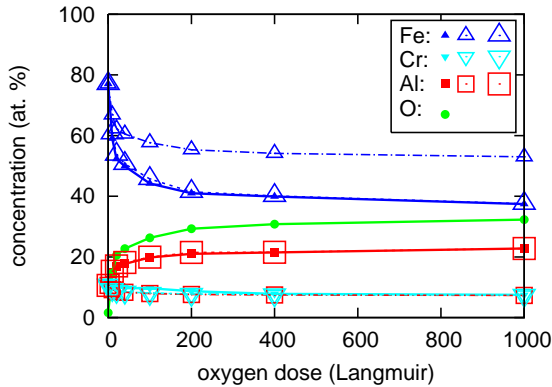}
\caption{The surface of Fe10Cr10Al at  700 $^{\rm o}$C, curves and symbols as explained in Fig.~\ref{fig1}.} \label{fig4}
\end{figure}

The main features of the FCR and VCR models are visualized in Fig.\ \ref{f_schema}. The analysis is made by optimizing the $k$ parameters to give the best fit to the experimental data. The concentrations 
\begin{eqnarray} 
c_{\rm X}(\theta)&=&100(1+k_{\rm X}(\theta))N_{\rm X}(0)/[(1+k_{\rm Fe}(\theta))N_{\rm Fe}(0)\nonumber \\
&+&(1+k_{\rm Cr}(\theta))N_{\rm Cr}(0)\nonumber \\
&+&(1+k_{\rm Al}(\theta))N_{\rm Al}(0)+N_{\rm O}(\theta)]\label{e_con}
\end{eqnarray} 
corresponding to the optimal $k$ parameters are shown in Figs.\ \ref{fig1}--\ref{fig4}. The number of oxygen atoms $N_{\rm O}$ in Eq. (\ref{e_con}) is taken from the experiments. 

%\section{Results and Discussion}
Figures~\ref{fig1}--\ref{fig4} show the experimental data (solid curve) and the predictions of the FCR (chain curve, small symbols) and VCR (dashed curve, large symbols) models. At room temperature both models predict the concentrations quite well. The good result of FCR (with zero $k$ parameters) indicates that in room temperature there exists no significant long range metal atom diffusion. 

The situation is changed dramatically when the temperature is raised up to 700 $^{\rm o}$C. 
As Figures \ref{fig2} and \ref{fig4} show the FCR model fails in the Fe and Al concentrations for both alloys. Contrary to Fe and Al, the Cr concentration (Fig. \ref{fig4}) seems to be quite well predicted by the FCR model. This is in accordance with the low integrated inter-diffusion constant of Cr in (Fe,Cr)$_3$Al$_2$. \cite{tortorici_1998} The optimal VCR result, shown in Figs.\ \ref{fig1}--\ref{fig4}, turns out to be within the Fe-Al exchange regime, i.e. 
\begin{equation} 
k_{\rm Fe}(\theta)N_{\rm Fe}(0)\equiv-k_{\rm Al}(\theta)N_{\rm Al}(0), \quad 
k_{\rm Cr}(\theta)\equiv 0. \label{e_ex} 
\end{equation} 
The optimized VCR $k$ parameters are shown in Table \ref{t_kparam}. Renormalizing with the bulk concentrations the initial surface enrichment of Al ($k_{\rm Al}(10)$) at 700 $^{\rm o}$C of Fe10Cr10Al is about 30~\% higher than that of Fe13Al.
\begin{table}[h]
\centering
\caption{\label{t_kparam}Optimized (best fit to experiment) VCR $k$ parameters as a function of oxygen dose ($\theta$).}
\begin{tabular}{ccccccccccc} 
\hline
\hline
& \multicolumn{4}{c}{room temperature}& \multicolumn{4}{c}{700 $^{\rm o}C$}\\
\hline
& \multicolumn{2}{c}{Fe13Al}& \multicolumn{2}{c}{Fe10Cr10Al}& \multicolumn{2}{c}{Fe13Al}& \multicolumn{2}{c}{Fe10Cr10Al}\\
$\theta$& Fe& Al& Fe& Al& Fe& Al& Fe& Al\\
\hline
  10& $-0.01$& 0.05& $-0.02$& 0.14& $-0.08$& 0.65& $-0.09$& 0.66\\
 100& $-0.02$& 0.13& $-0.03$& 0.24& $-0.17$& 1.39& $-0.21$& 1.51\\
1000& $-0.02$& 0.20& $-0.04$& 0.28& $-0.34$& 2.78& $-0.29$& 2.12\\
\hline
\hline
\end{tabular}
\end{table} 
%\begin{table}[h]
%\centering
%\caption{\label{t_numb}Surface concentration (at. \%) according to XPS measurements.}
%\begin{tabular}{cccccc} 
%\hline
%\hline
%& \multicolumn{2}{c}{Fe13Al}& \multicolumn{3}{c}{Fe10Cr10Al}\\
%$\theta$& Fe& Al& Fe& Al& Cr\\
%\hline
%   0& 90& 10& 78& 12& 10\\
%\hline
%\hline
%\end{tabular}
%\end{table} 
%\begin{table}[h]
%\centering
%\caption{\label{t_ratio}Best fit of the total number of Fe-Al pair exchanges at 700 $^{\rm o}$C as a function of oxygen dose ($\theta$) according to the VCR model. Pair exchanges are expressed as per cents of the total number of Fe atoms in the surface region before the oxidation.}
%\begin{tabular}{ccc} 
%\hline
%\hline
%&\multicolumn{2}{c}{Fe-Al pair exchanges (\%)}\\
%\hline
%$\theta$& Fe13Al& Fe10Cr10AlCr\\
%\hline
%  10& 8& 9\\
% 100& 17& 21\\
%1000& 34& 29\\
%\hline
%\hline
%\end{tabular}
%\end{table} 

As Figs.~\ref{fig1}--\ref{fig4} show the initial oxidation of Fe10Cr10Al is faster than that of Fe13Al which can be related to the higher Al to Fe ratio of Fe10Cr10Al surface. Clearly Cr boosts the Al surface enrichment leading to faster growth of the protective oxide scale on Fe10Cr10Al. However, this shows up later as reduced oxidation rate because the oxide scale in Fe10Cr10Al has already reached the critical thickness of the protective scale. On the other hand, the Al-oxide scale in Fe13Al seems to be incomplete leading to faster oxidation at practical time scales and finally vulnerability in high temperature applications. 

To get more insight to the oxidation of Fe-Al and Fe-Cr-Al we compare the present results with our previous findings on the same subject. The basic atomic phenomena on the backstage of the improved oxidation resistance in Cr-containing Fe-Al alloys can be connected to the local environment induced changes in chemical potentials. Theoretical investigations show that Cr has a potential to activate a kind of Al-pump from the bulk to the surface. \cite{airiskallio_2010a} At 900 K the Al concentration at the surface of Fe5Al is increased from 63 to 72 (81) at.\% when 5 (10) at.\% Cr is added to the alloy. \cite{airiskallio_2010b} This theoretical prediction is in good agreement with the 30~\% increase of the surface enrichment of Al, as indicated by the present XPS data. 

\begin{figure}[h]
\includegraphics[width=0.5\textwidth,angle=0]{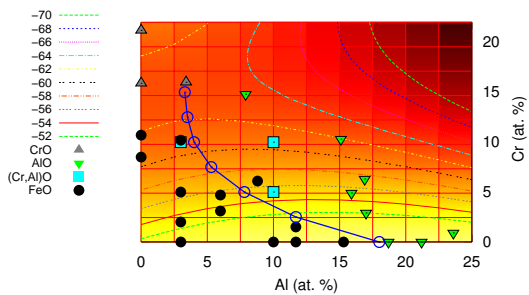}
\caption{Calculated chemical potential difference ($\Delta E$ in mRy) as a function of Cr and Al concentrations in bulk. Experimental surface oxide types above 1000~$^{\rm o}$C (symbols). \cite{tomaszewicz_1978}} \label{fig_map}
\end{figure}

Figure\ \ref{fig_map} shows the calculated chemical potential difference \cite{airiskallio_2010b} $\Delta E = (\mu_{\rm Fe}-\mu_{\rm Al})^{\rm bulk}-(\mu_{\rm Fe}-\mu_{\rm Al})^{\rm surface}$ and the measured surface oxide type. \cite{tomaszewicz_1978} A curve (open symbols) shows theoretical bulk compositions leading to the same high-temperature Al surface concentrations as the Fe18Al alloy. Surprisingly, this calculated curve separates the good (Al-O) and poor (Fe-O) oxidation resistance regions quite well suggesting that the Al surface concentration has an important effect on the high temperature oxidation resistance. Therefore, the effect of Cr on the high temperature oxidation resistance seems to be indirect via the change of the chemical potentials. The dividing curve in Fig.\ \ref{fig_map} is obtained iteratively from the equation
\begin{equation}
c_{\rm Al}^{\rm b}= 18e^{-(\Delta E(c_{\rm Al}^{\rm b},c_{\rm Cr}^{\rm b})-\Delta E(18,0))/kT},
\end{equation}
where an exponential acceptance criterion for an atomic jump is adopted.

%\section{Conclusions}
In summary, excellent reproduction of the experimental surface concentrations during the initial oxidation of Fe13Al and Fe10Cr10Al can be obtained by a model of Cr boosted exchange type diffusion of Fe and Al. The present XPS measurements give direct evidence of the accelerated alumina growth on Fe10Cr10Al. 10~\% Cr in Fe-Al increases the Al driving force to the surface by about 30~\% in initial oxidation. 
%\begin{acknowledgements} 
%\end{acknowledgements} 

%\bibliography{citations}

\end{document}